# Tomographic Phase Imaging with Randomized Probe Imaging

**JORDAN T. O'NEAL,[1,*] BYEONGDU LEE,[1] SOENKE SEIFERT,[1] HANNA RUTH,[1] MICHAEL WOJCIK,[1] HAIDAN WEN,[1,2] YANQI LUO,[1] YI JIANG,[1] AND JUNJING DENG[1]**

[1]*Advanced Photon Source, Argonne National Laboratory, Lemont, IL 60439, USA.*
[2]*Material Science Division, Argonne National Laboratory, Lemont, IL 60439, USA.*
**oneal@anl.gov*

**Abstract:** We demonstrate tomographic phase imaging of cubic gold nanoparticles with sub-100 nm resolution requiring just a single far-field coherent diffraction pattern per projection. By using randomized probe imaging (RPI), a real-space amplitude and phase image can be reconstructed from a single diffraction pattern. These phase images are then fed into a standard tomographic workflow to retrieve the 3D volume. Nanoscale X-ray phase tomography has often relied on ptychography, which requires slow 2D scanning for each projection. By using RPI, the data collection is greatly sped up at the cost of some resolution. We compare an RPI-tomography volume to a ptychography-tomography volume of the same sample, finding high consistency between the two methods. To the best of our knowledge, this is the first application of RPI to tomography.

## 1. Introduction

Quantitative three-dimensional phase imaging at the nanoscale enables the study of the internal structure and composition of materials, especially those with low absorption contrast. There is interest in collecting tomograms fast enough for *in-situ* time-resolved measurements [1–5], but nanoscale resolution phase tomography is generally slow, often taking several minutes for low signal-to-noise-ratio (SNR) tomograms and longer for good statistics [6]. Achieving phase tomography in a few seconds with approximately 100 nm resolution would enable *in-situ* measurements of crystal formation [3], fuel cell operation [4], and foam compression [5].

Phase nano-tomography is generally done by gathering real-space 2D projections at many angles and then reconstructing them into a 3D volume. Several imaging methods can be used to supply these two-dimensional projections which balance speed, imaging area, radiation dose, and resolution. Full-field Zernike phase contrast imaging can image large volumes at moderate resolution [7] or at high speed [8], but it requires inefficient optics after the sample, increasing the radiation dose on the sample. Ptychography is a coherent diffractive imaging technique which can provide the highest resolution phase imaging with efficient use of the radiation dose, but the required spatial overlap during area scanning increases the data volume and imaging time [3,9]. Holography achieves good resolution and is faster than ptychography, but it still requires three to four images per projection [10]. A version of holography was recently proposed which uses coded apertures and a joint reconstruction algorithm to achieve tomographic reconstruction with just one near-field diffraction pattern per projection with moderate resolution of 50-100 nm [11]. However, this method requires synchronized motion of the sample and coded aperture, and the proposed algorithm assumes a completely known coded aperture and probe function.

We demonstrate the use of randomized probe imaging (RPI) [12] to achieve tomographic phase image reconstruction from a single far-field diffraction pattern per projection. In RPI, the sample is illuminated with a structured probe created by a randomized zone plate (RZP). By using an initial probe taken from a calibration ptychography scan, an iterative algorithm can be used to reconstruct the sample amplitude and phase. The scheme incorporates a bandwidth-limiting constraint into a standard ptychography algorithm [12], allowing for stable reconstructions with just a single diffraction pattern. The bandwidth-limiting constraint effectively enlarges the reconstructed pixel size, and the half-pitch resolution is limited to the outer zone width of the RZP under optimal conditions. The individual projections can then be used in a standard tomography workflow to retrieve a 3D volume.

In this work we reconstruct a 3D tomographic volume of a heterogeneous cluster of approximately 100 nm cubic gold nanoparticles (AuCs) from a set of far-field diffraction patterns, one for each projection. In the first step, we use RPI to retrieve real-space 2D projections from each diffraction pattern separately. These 2D projections are then used in a standard tomographic reconstruction. The diffraction patterns used for RPI-tomography were taken from a larger ptychography-tomography dataset. We compare the tomography volumes resulting from the RPI-derived projections and the ptychography-derived projections, validating the robustness of RPI-tomography.

## 2. Methods

Measurements were performed at the 12-ID-E SAXS beamline of the upgraded Advanced Photon Source. Coherent X-rays at 8 keV were focused by a RZP with an outer zone width of 50 nm and a diameter of 250 µm. The optic created a 7 µm diameter focal spot size consisting of random speckles with approximately 50 nm size, forming structured illumination suitable for RPI, shown in Figure 1 (e). The sample, a cluster of heterogeneous AuCs dispersed on a silicon nitride membrane, was placed near the focal plane in a few-centimeter airgap. The diffracted light passed into an evacuated flight tube and was measured by a Dectris Pilatus 2M photon counting detector located 10 m from the sample.

The RZP pattern was designed following the methodology laid out by Levitan, *et. al.* [12,13]. The spot size was chosen to fill 80% of the field of view, which allows a larger imaging area but reduces the maximum resolution achievable with RPI. The RZP was fabricated at the Center for Nanoscale Materials by modifying a method previously published [14]. Starting with a silicon nitride membrane coated with Ti/Au, a JEOL JBX-8100FS electron beam lithography tool was used to pattern a 500 nm thick poly (methyl methacrylate) (PMMA) positive resist with the designed RZP pattern. After development, the remaining resist acted as an electroplating mold which was then filled with Au. The resist mold was then removed using a chemical solvent and the RZP pattern was prepared for the X-ray experiment.

AuCs were synthesized with seed-mediated growth according to published methods [15]. Cube sizes (110.5 ± 8.6 nm, 88.0 ± 4.8 nm, and 67.2 ± 3.9 nm) were varied by changing the volume of seed particle solution added to the synthesis. AuCs were functionalized with 3' thiol-terminated oligonucleotides (TCA ACT ATC TCT AGC TAC Sp182 SH). Briefly, 10 nmol of DNA was added per mL of synthesis and 0.1 M NaCl increments were added per hour up to a final concentration of 0.4 M NaCl. The particles were kept on a shaker at 900 rpm at room temperature during salt additions and left overnight. Then, DNA-functionalized AuCs were washed 3 times in DI water, concentrated to 200 uL in 0.3 M NaCl, and 20 nmol of linker DNA (GTA GCT AGA GAT AGT TGA Sp18 GTACGCG) was added to the solution. This solution was aliquoted into PCR tubes and slow-cooled in a thermal cycler from 75 – 25 °C at a rate of 0.1 °C / 20 min. After assembly, cubic colloidal crystals were silica embedded according to previous methods [16] to stabilize the samples and were drop-cast onto substrates for electron microscopy and X-ray imaging.

The initial probe function was characterized by measuring a Siemens star with ptychography, using a step size of 500 nm. In ptychography, diffraction patterns are gathered by scanning the sample with a coherent beam across many partially overlapping positions. An iterative phase retrieval algorithm reconstructs the complex sample and probe simultaneously. For the ptychography results, we use the least-squares maximum likelihood algorithm implemented in the *Pty-Chi* software package [17], using 10 probe modes and 3 orthogonal probe relaxation (OPR) modes [18]. After calibration, we measured a ptychography-tomography scan of the AuC sample. The beam was scanned in 750 nm steps in a regular grid pattern over an area of 10×8 μm$^2$ with an exposure time of 2 ms. The average imaging rate was 2.5 Hz, limited by the settling time between step scan positions and overhead between angle scans, and the entire scan took approximately four hours. The ptychography results have a pixel size of 18 nm, but the resolution was limited by mechanical and beam stability to approximately 30 nm half-pitch.

The RPI scan was assembled by selecting one diffraction pattern from each ptychography projection. From a single diffraction pattern, a real-space image was reconstructed using the RPI method in the CDTools software package [19], using the calibrated probe from the initial scan dropping the OPR modes. The diffraction pattern was cropped to 512×512 pixels, and the bandwidth-limiting constraint was used to retrieve a 200×200 pixel image. The L-BFGS method with mild regularization was done for 20 iterations with the object array having two incoherent modes, followed by 60 iterations with a single mode. Allowing for an incoherent object at the onset of iterative phase retrieval has been found to aid the convergence [13].

The tomography scan was taken from -50° to +60° in 0.5° steps. We were unable to measure a full 180° due to the edges of the silicon chip, degrading the depth resolution in the final result due to the missing wedge issue. The Crowther criterion gives a maximum achievable half-pitch resolution of 30 nm, and the 70° missing wedge should degrade the longitudinal resolution by a factor of 1.7 [20]. The vertical and horizontal displacement error of each projection was corrected using the GUI-based workflow in the *pyxalign* software package [21]. Each of the tomographic datasets were aligned by first using cross-correlation to correct the jitter between angularly adjacent projections and then using multi-resolution projection-matching alignment to retrieve the final high-precision alignment. The 3D reconstructions of both the RPI- and ptychography-derived projections were obtained by seeding a simultaneous algebraic reconstruction technique (SART) [22] reconstruction with the result of a filtered back-projection (FBP) reconstruction.

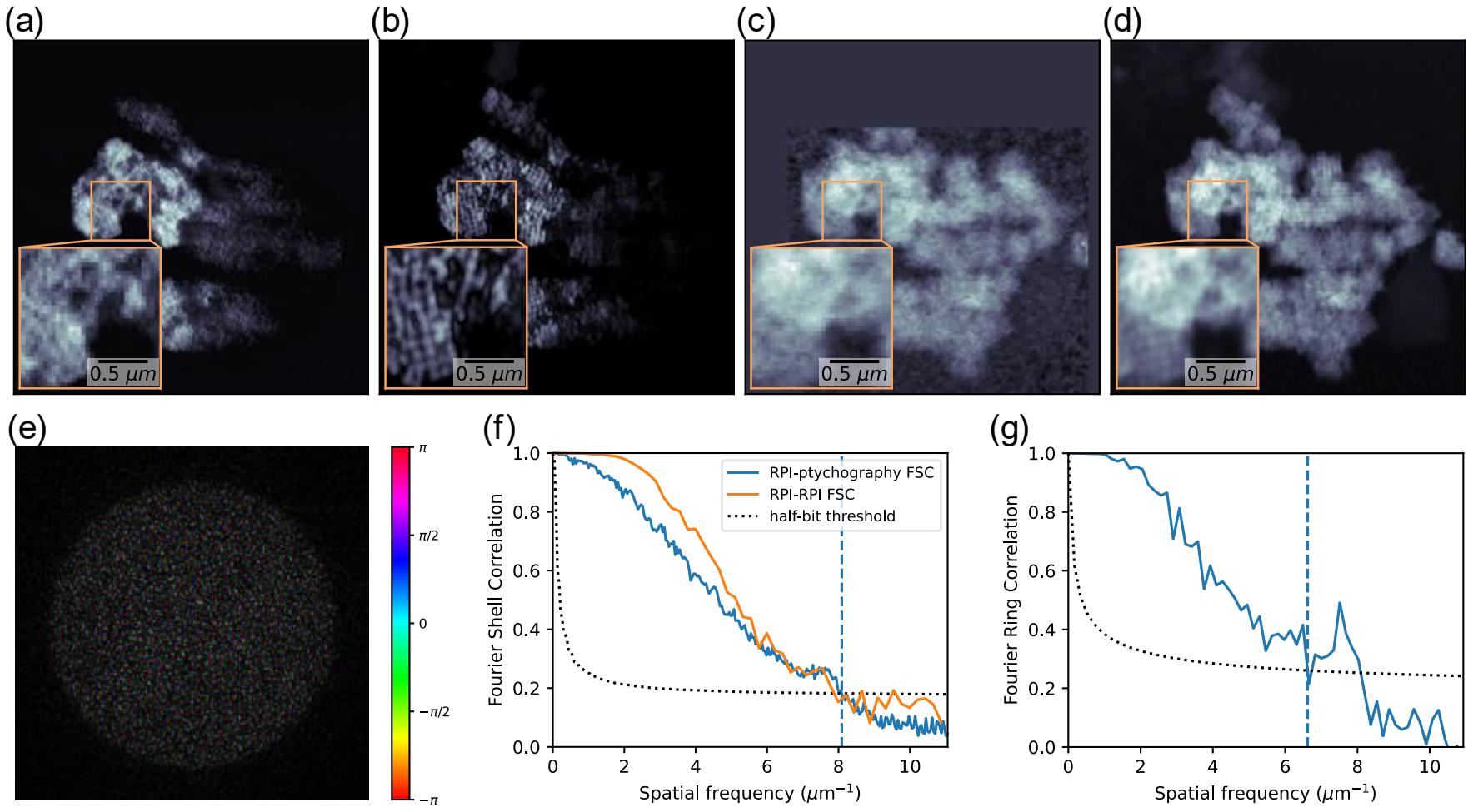

Fig. 1. Representative slice of (a) RPI-tomography and (b) ptychography-tomography reconstructions. (c, d) Projections at 0° retrieved by RPI and ptychography, respectively. (e) Complex initial probe retrieved from ptychography shown in HSV colorscale. (f) Fourier shell correlations between RPI- and ptychography-tomography, and also splitting the RPI dataset in two. (g) Fourier ring correlation between RPI and ptychography projections at 0°. All correlations use the half-bit criteria.

## 3. Results

Fig. 1 summarizes the RPI-tomography reconstruction. Representative slices of the RPI-tomography and ptychography-tomography volumes are shown in panels (a) and (b), respectively, sliced along the *x*-*y* plane. For comparison, the 0° RPI (c) and ptychography (d) projections are shown, which also view the *x*-*y* plane. In both ptychography results, individual

AuCs can be resolved, while RPI was unable to resolve them. Despite the difference in resolution, the larger features match well between RPI and ptychography.

To quantify the resolution and test for artifacts introduced by RPI, a Fourier shell correlation (FSC) was calculated between the RPI- and ptychography-tomography volumes (f). Using the half-bit criterion, the half-pitch 3D resolution was found to be 62 nm, slightly too large to resolve individual AuCs with side length approximately 100 nm, and too large to resolve the small gaps between nanoparticles. As confirmation, an FSC was computed by splitting the RPI projections and running FBP on each half (f). Panel (g) shows a 2D Fourier ring correlation (FRC) between the projections in panels (c) and (d), showing a resolution of 76 nm. Within the noise of the FRC curve, the resolution is similar between the projections and retrieved volumes, despite the missing angles. This could be a sign that RPI-tomography has the potential to overcome the resolution limit imposed by the Nyquist frequency in single-frame RPI by integrating information between multiple diffraction patterns in the tomography reconstruction. In this work, we used existing tools to perform a simple, two-step reconstruction. Algorithmic improvements are likely possible which reconstruct the 3D volumes directly from the diffraction patterns.

The data shown here were acquired over four hours with a ptychography-tomography scan. However, the RPI-tomography data required just 220 diffraction patterns. An angle step scan with our system can be performed at 3 Hz, requiring 73 s for the entire scan, a factor of 200 improvement. Furthermore, fly scanning at our maximum detector rate of 30 Hz could complete the same scan in 7.3 s, and a faster detector could shorten this time further. The ultimate limit is the available X-ray flux and the blurring induced by an angular fly scan. The currently available speed compares favorably to full-field X-ray phase tomography [6], where in one case 15 minutes was required to match our resolution, and a 3 minute scan was degraded to 82 nm half-pitch, with a similar field of view.

## 4. Conclusion

We demonstrate that RPI-tomography has the potential to provide nanoscale-resolution 3D phase imaging which requires just a single far-field diffraction pattern for each projection. Despite the small angular range of 110°, tomographic reconstruction was successful and had a resolution comparable to that of the 2D projections. Comparison with a standard ptychography-tomography reconstruction of the same sample shows the reliability of the RPI-tomography method. The result shown here was assembled by taking one diffraction pattern from each ptychography scan, but the measurement process could be greatly shortened to one minute for an angular step scan or shorter with an angular fly scan. RPI-tomography holds potential for time-resolved tomography of nanoscale dynamics such as crystal formation [3] and fuel cell operation [4], and further development is necessary to unlock its full potential.

## 5. Back matter

**Funding.** U.S. DOE Office of Science-Basic Energy Sciences, under Contract No. DE-AC02-06CH11357.

**Acknowledgment.** We thank Rachel Chan for providing us with cubic gold nanoparticle samples. Work performed at Advanced Photon Source on APS beam time award(s) (DOI: https://doi.org/10.46936/APS-191625/60015317) and the Center for Nanoscale Materials, both U.S. Department of Energy Office of Science User Facilities, and is based on work supported by Laboratory Directed Research and Development (LDRD) funding from Argonne National Laboratory, provided by the Director, Office of Science, of the U.S. DOE under Contract No. DE-AC02-06CH11357.

**Data availability.** Data underlying the results presented in this paper are not publicly available at this time but may be obtained from the authors upon reasonable request.

## Supplementary Material

This supplementary material contains additional views of the tomographic reconstructions and the translations used in the multi-resolution projection-matching alignment.

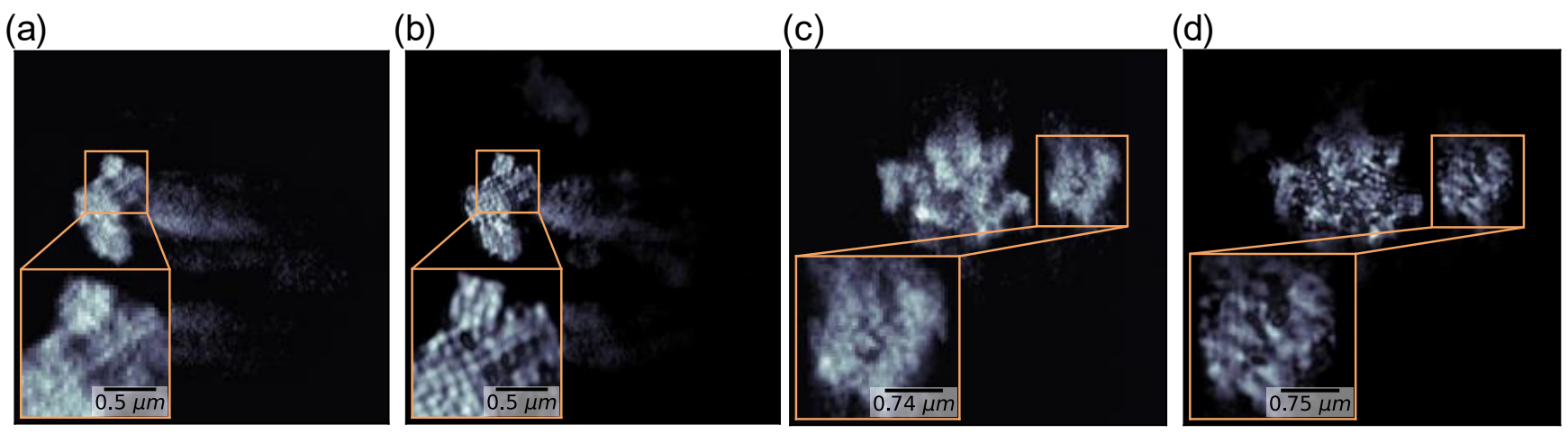

Fig S1. Additional slices of the RPI-tomography (a, c) and ptychography-tomography (b, d) volumes. (a, b) Slice along the *x-y* plane. (c, d) Slice along the *y-z* plane.

Fig. S1 shows additional slices of the tomography volumes. Panels (a) and (b) show a slice along the same x-y plane shown in Fig. 1, but offset by approximately 630 nm. Slices offset by a large *z* distance appear well-separated without any apparent cross-talk. However, the limited depth resolution can be seen in panels (c) and (d) which show a slice along the y-z plane.

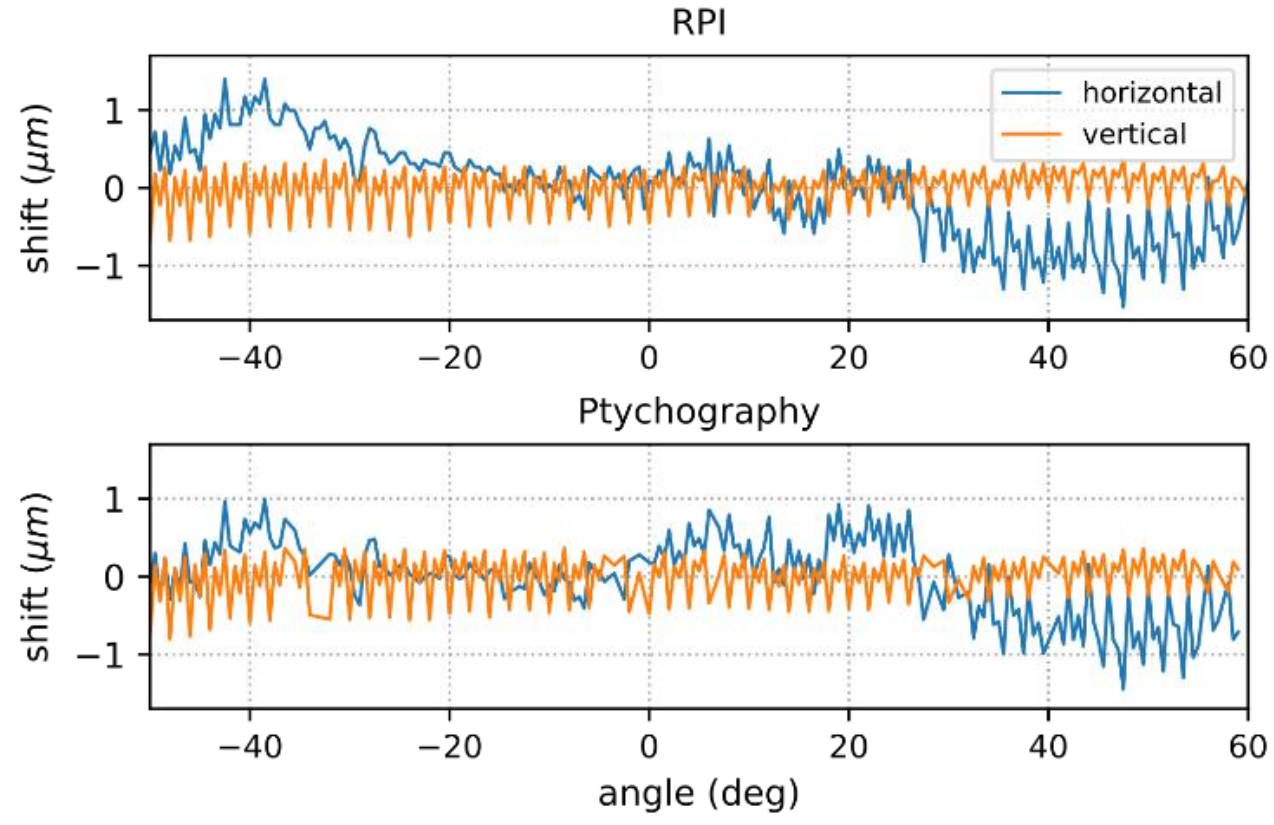

Fig. S2. Total shift displacements found with projection-matching alignment for both tomography reconstructions.

Fig. S2 shows the total shift displacements applied to each projection before tomography reconstruction. The angular scans were performed in 2° steps, with four sets of rotations offset by 0.5°. This appears as a period-four cycle in the shift displacements. Even after four hours of scanning, the displacements were within 1-2 µm, and so fast RPI scanning should be possible with a high degree of alignment.